\documentstyle[12pt,here]{article}

\catcode`\@=11
\long\def\@makefntext#1{ 
\protect\noindent \hbox to 3.2pt {\hskip-.9pt
$^{{\ninerm\@thefnmark}}$\hfil}#1\hfill} 

 \def\@makefnmark{\hbox to 0pt{$^{\@thefnmark}$\hss}}  

\def\ps@myheadings{\let\@mkboth\@gobbletwo
\def\@oddhead{\hbox{} 
\rightmark\hfil\ninerm\thepage}
\def\@oddfoot{}\def\@evenhead{\ninerm\thepage\hfil 
\leftmark\hbox{}}\def\@evenfoot{}
\def\sectionmark##1{}\def\subsectionmark##1{}}

\newcounter{sectionc}\newcounter{subsectionc}\newcounter{subsubsectionc}
\renewcommand{\section}[1] {\vspace{0.6cm}\addtocounter{sectionc}{1}
\setcounter{subsectionc}{0}\setcounter{subsubsectionc}{0}\noindent
	{\bf\thesectionc. #1}\par\vspace{0.4cm}}
\renewcommand{\subsection}[1] {\vspace{0.6cm}\addtocounter{subsectionc}{1}
	\setcounter{subsubsectionc}{0}\noindent
	{\it\thesectionc.\thesubsectionc. #1}\par\vspace{0.4cm}}
\renewcommand{\subsubsection}[1]
{\vspace{0.6cm}\addtocounter{subsubsectionc}{1}
	\noindent {\rm\thesectionc.\thesubsectionc.\thesubsubsectionc.
	#1}\par\vspace{0.4cm}}

\newcounter{appendixc}
\newcounter{subappendixc}[appendixc]
\newcounter{subsubappendixc}[subappendixc]

\renewcommand{\appendix}[1] {\vspace{0.6cm}
        \refstepcounter{appendixc}
        \setcounter{figure}{0}
        \setcounter{table}{0}
        \setcounter{equation}{0}
        \renewcommand{\thefigure}{\Alph{appendixc}.\arabic{figure}}
        \renewcommand{\thetable}{\Alph{appendixc}.\arabic{table}}
        \renewcommand{\theappendixc}{\Alph{appendixc}}
        \renewcommand{\theequation}{\Alph{appendixc}.\arabic{equation}}
        \noindent{\bf Appendix \theappendixc #1}\par\vspace{0.4cm}}

\def\abstracts#1{{
	\centering{\begin{minipage}{30pc}\tenrm\baselineskip=12pt\noindent
	\centerline{\tenrm ABSTRACT}\vspace{0.3cm}
	\parindent=0pt #1
	\end{minipage} }\par}}

\renewenvironment{thebibliography}[1]
	{\begin{list}{\arabic{enumi}.}
	{\usecounter{enumi}\setlength{\parsep}{0pt}
\setlength{\leftmargin 1.25cm}{\rightmargin 0pt}
	 \setlength{\itemsep}{0pt} \settowidth
	{\labelwidth}{#1.}\sloppy}}{\end{list}}

\newcounter{itemlistc}
\newcounter{romanlistc}
\newcounter{alphlistc}
\newcounter{arabiclistc}

\newcommand{\fcaption}[1]{
        \refstepcounter{figure}
        \setbox\@tempboxa = \hbox{\tenrm Fig.~\thefigure. #1}
        \ifdim \wd\@tempboxa > 6in
           {\begin{center}
        \parbox{6in}{\tenrm\baselineskip=12pt Fig.~\thefigure. #1 }
            \end{center}}
        \else
             {\begin{center}
             {\tenrm Fig.~\thefigure. #1}
              \end{center}}
        \fi}

\newcommand{\tcaption}[1]{
        \refstepcounter{table}
        \setbox\@tempboxa = \hbox{\tenrm Table~\thetable. #1}
        \ifdim \wd\@tempboxa > 6in
           {\begin{center}
        \parbox{6in}{\tenrm\baselineskip=12pt Table~\thetable. #1 }
            \end{center}}
        \else
             {\begin{center}
             {\tenrm Table~\thetable. #1}
              \end{center}}
        \fi}

\def\@citex[#1]#2{\if@filesw\immediate\write\@auxout
	{\string\citation{#2}}\fi
\def\@citea{}\@cite{\@for\@citeb:=#2\do
	{\@citea\def\@citea{,}\@ifundefined
	{b@\@citeb}{{\bf ?}\@warning
	{Citation `\@citeb' on page \thepage \space undefined}}
	{\csname b@\@citeb\endcsname}}}{#1}}

\def\@cite#1#2{\unskip\nobreak\relax
    \def\@tempa{$\m@th^{\hbox{\the\scriptfont0 #1}}$}%
    \futurelet\@tempc\@citexx}
\def\@citexx{\ifx.\@tempc\let\@tempd=\@citepunct\else
    \ifx,\@tempc\let\@tempd=\@citepunct\else
    \let\@tempd=\@tempa\fi\fi\@tempd}
\def\@citepunct{\@tempc\edef\@sf{\spacefactor=\the\spacefactor\relax}\@tempa
    \@sf\@gobble}

\def\citenum#1{{\def\@cite##1##2{##1}\cite{#1}}}
\def\citea#1{\@cite{#1}{}}

\newcount\@tempcntc
\def\@citex[#1]#2{\if@filesw\immediate\write\@auxout{\string\citation{#2}}\fi
  \@tempcnta\z@\@tempcntb\m@ne\def\@citea{}\@cite{\@for\@citeb:=#2\do
    {\@ifundefined
       {b@\@citeb}{\@citeo\@tempcntb\m@ne\@citea\def\@citea{,}{\bf ?}\@warning
       {Citation `\@citeb' on page \thepage \space undefined}}%
    {\setbox\z@\hbox{\global\@tempcntc0\csname b@\@citeb\endcsname\relax}%
     \ifnum\@tempcntc=\z@ \@citeo\@tempcntb\m@ne
       \@citea\def\@citea{,}\hbox{\csname b@\@citeb\endcsname}%
     \else
      \advance\@tempcntb\@ne
      \ifnum\@tempcntb=\@tempcntc
      \else\advance\@tempcntb\m@ne\@citeo
      \@tempcnta\@tempcntc\@tempcntb\@tempcntc\fi\fi}}\@citeo}{#1}}
\def\@citeo{\ifnum\@tempcnta>\@tempcntb\else\@citea\def\@citea{,}%
  \ifnum\@tempcnta=\@tempcntb\the\@tempcnta\else
   {\advance\@tempcnta\@ne\ifnum\@tempcnta=\@tempcntb \else \def\@citea{--}\fi
    \advance\@tempcnta\m@ne\the\@tempcnta\@citea\the\@tempcntb}\fi\fi}

\def\fnt#1#2{\footnotetext{\kern-.3em
	{$^{\mbox{\sevenrm #1}}$}{#2}}}

 1
 1
 1

\font\tenbf=cmbx10
\font\tenrm=cmr10
\font\tenit=cmti10

\font\ninerm=cmr9

\begin{document}
\begin{center}
\begin{flushright}
TIFR/TH/95-04
\end{flushright}
{\tenbf RESONANCES IN SMALL FERMI SYSTEMS} \\
\vspace{0.8cm}
{\tenrm MUSTANSIR BARMA and R.S. BHALERAO}\\
\baselineskip=13pt
{\tenit Theoretical Physics Group, Tata Institute of Fundamental
Research,}\\
\baselineskip=12pt
{\tenit Homi Bhabha Road, Colaba, Bombay 400 005, India}
\bigskip
\end{center}
\vspace{0.9cm}
\abstracts{
The systematics of the size dependence of the resonant response of
small metal particles and nuclei to incident electromagnetic radiation
is studied. The known radius$^{-1}$ variation of the full width at
half maximum (FWHM) in matrix-embedded metal particles is
qualitatively accounted for by a quantum calculation of the response
within a simple model. In free clusters, the behaviour is more
complicated, possibly because of thermal excitation of surface
modes. For nuclei, the FWHM shows strong shell-structure-linked
oscillations across the periodic table. Focussing on the lower
envelope of the oscillations (magic nuclei), the downward trend of the
FWHM is consistent with the radius$^{-1}$ variation. A schematic
theoretical description of the systematics in nuclei is presented. If
the FWHMs are scaled by the respective Fermi energies and the inverse
radii by the Fermi wave vectors, the data sets for matrix-embedded
metal particles and nuclei become comparable in magnitude.
}

\vskip 1cm
\setlength{\baselineskip}{14pt}
\noindent {\bf 1. Introduction}
\vskip 0.5cm

The resonant response of small Fermi systems to electromagnetic
radiation has been receiving a good deal of attention in various
branches of physics. By `small' we mean a system whose dimensions are
smaller than, or at most comparable to, the mean free path of the
fermions and also the wavelength of the incident radiation. Two sets
of physical systems which satisfy these criteria are small metal
particles and clusters on the one hand, and atomic nuclei on the
other. In this contribution, we shall compare the systematics of
resonances in these two sets of systems, namely the Mie resonance in
metal particles and the giant dipole resonance (GDR) in nuclei, with
varying numbers of fermions.

Certain points of resemblance between the electronic properties of
metal particles and the properties of nuclei have been discussed
earlier, see, {\it e.g.}, Sugano \cite{sugano}. Electrons in metal
particles and nucleons in nuclei both constitute finite Fermi systems
at temperatures much less than the respective Fermi energies -- a fact
which cuts across the very different scales of length, mass and energy
in the two systems. Metal particles, like nuclei (and unlike atoms),
exhibit saturation, or constancy of particle density, with increasing
size. Also, shell structure in energy levels -- long familiar in
nuclear physics -- manifests itself in the relative abundances,
polarizabilities and ionization potentials of metal clusters as well
\cite{deHeer}, although the large magic numbers in the two cases differ
because of the different strengths of the spin-orbit force.

\vskip 0.5cm
\hrule width 7 cm
\vskip 0.1cm
\noindent To appear in {\it Cooperative Effects in Many-Electron Systems
and their Response to External Fields}, edited by J.-P. Connerade,
(World Scientific, Singapore).
\newpage

Here we focus on the analogy between the response of metal particles
and nuclei to electromagnetic radiation. In metal particles, the Mie
resonance involves displacing the conduction electron cloud with
respect to the background of positive ions, and there are
electromagnetic restoring forces. \cite{bw,apell,kresin} In a nuclear
giant dipole resonance, protons are displaced with respect to neutrons
and strong interactions provide the restoring force. \cite{bf,vdw}

In Section 2, we discuss the Mie resonance in metal particles with an
emphasis on size dependence of the resonance width.  We also present
there a calculation of the width within the framework of linear
response theory. In Section 3, we discuss cross section systematics of
GDR in nuclei and the various contributions to its width. In Section
4, we point out similarities and differences between the systematics
in metal particles and nuclei. Conclusions are given in Section 5.

Some of the results discussed in this paper have appeared elsewhere.
\cite{bb}

\vskip 1cm
\noindent {\bf 2. The Mie Resonance in Small Metal Particles}
\vskip 0.5cm

The absorption spectrum of small metal particles shows a strong
resonance in the optical region. Studied by G. Mie in the early part
of the century, the origin of this resonance can be understood from
the following simple argument \cite{bw}, in the limit that the
wavelength $\lambda$ of the radiation far exceeds the radius $R$ of
the particle.  Recalling that the field that penetrates a sphere of
dielectric constant $\varepsilon$ in a uniform external electric
field $E_o$ is
$3 E_o/(2 + \varepsilon)$, we see that there is resonant penetration
at the frequency $\omega_0$ at which $\varepsilon(\omega_0) = -2$.
Using the free electron gas expression
$\varepsilon(\omega)=1-\omega_p^2/\omega^2$, the resonance frequency
is given by $\omega_0=\omega_p/ \sqrt {3}$, where $\omega_p
=\sqrt{4\pi n e^2/m}$ is the plasma frequency, $n$ being the electron
number density. Thus, when electromagnetic radiation of frequency
$\omega_0$ falls on a metal particle with radius $R << \lambda,$ there
is a resonant response.

The systematics of the Mie resonance in the regime $R << \lambda$ have
been studied experimentally for two types of samples: (i) Metal
particles embedded in various matrices such as glass and the inert
element solids. In such samples, particle agglomeration can pose a
problem, but it is sometimes possible to achieve mutual isolation of
particles.  However, a spread of sizes cannot be avoided, a typical
spread of radius being $\sim 20\%$.  (ii) Free metal clusters in a
beam, separated size-wise by mass spectroscopy.  Measurements of the
photoresponse are done while the particles are in beam, and often
quite hot, though it is difficult to pin down the temperature
reliably.

In Section 2.1 below, we briefly summarize the results of experiments
 on type (i) samples, and outline a linear response RPA calculation of
 the width. An inverse-size dependence is obtained, and the
 proportionality constant is calculated, correcting a long-standing
 error. Experiments on type (ii) samples are discussed in Section
 2.2. This is followed by a discussion of physical effects which may
 contribute to the observed differences of behaviour of samples of
 types (i) and (ii), in particular the role of surface fluctuations at
 finite temperature.
\newpage
\vskip 1cm
\noindent {\it 2.1. Matrix-held Particles: 1/R Law for the FWHM}
\vskip .5cm

A fairly large range of sizes has been investigated in experiments on
samples of type (i); for a review, see Ref. 9. It is found that
$\omega_0$ does not vary very strongly with size --- less than $10\%$
as the size is changed from $\sim 10 \AA$ to $\sim 100
\AA $. This is not surprising,
as $\omega_p$ and thus $\omega_0$ depends primarily only on the
electron density.  The sign of the shift of $\omega_0$ (towards the
red or blue end of the spectrum) has been a subject of debate.  But
more dramatic is the effect of size variation on the full width at
half maximum (FWHM), which changes several-fold. In fact, experimental
results reveal a systematic dependence of $\Gamma_{av}$ (FWHM averaged
over the spread of sizes in a sample) on $R$:
\begin{equation}
\Gamma_{av} = K ~{\hbar v_F \over R} + \Gamma_\infty,
\end{equation}
where $v_F$ is the Fermi velocity, $K$ is a constant of order unity
and $\Gamma_\infty$ is the width in the bulk medium. Equation (1)
describes the variation of the linewidth of $Ag$ particles in a
variety of host matrices. The constant $K$ depends on the matrix \cite{kg};
it goes down by a factor $\sim 3$ as the
matrix is changed from glass to an inert element solid like $Ar$ and
$Ne$, presumably due to differences in the potential at the outer
surface of each metal particle.

Calculation of the photoabsorption by small metallic spheres was first
attempted by Kawabata and Kubo \cite{kk} within a simple model; an
error in their calculation was corrected in Ref. (11). Below we
briefly recapitulate the principal steps of the calculation. Suppose a
volume fraction $\alpha$ of identical, small spherical particles with
dielectric constant $\varepsilon
\equiv \varepsilon_1 -i \varepsilon_2$ is well
dispersed in a medium with real dielectric constant
$\varepsilon_0$. The absorption coefficient is given by \cite {kk,bs}
\begin{equation}
\gamma =  { 18 \pi
\alpha \varepsilon_0^{3/2} \over \lambda }~{\varepsilon_2 \over
(2\varepsilon_0 + \varepsilon_1)^2 + \varepsilon_2^2},
\end{equation}
provided that $\lambda >> R$.  The Mie resonance occurs when $2
\varepsilon_0 + \varepsilon_1 = 0$. The lineshape is approximately a
Lorentzian centered at the resonance frequency $\omega_0$.

The random phase approximation (RPA) for the system involves
calculating the dielectric constant using a noninteracting electron
gas model \cite{kk}. The confining spherical potential well is taken
to be infinitely deep. Within linear response theory,
$\varepsilon_2(\omega)$ can be expressed in terms of the one-electron
energies $E_i$ and eigenfunctions $| i>$:
\begin{equation}
\varepsilon_2(\omega) = {4 \pi^2 \hbar \over
\omega^3 V} ~ \sum_{i,j} {f(E_i)(1-f(E_j)) \over E_j-E_i} ~ |< j| \dot
J_z  | i> |^2~ \delta ( \hbar \omega - E_j + E_i).
\end{equation}
Here $V$ is the volume of the spherical well, $\dot J_z$
is the rate of change of the current operator and $f(E_i)$ is the
occupation number of state i.

    In a hard-walled sphere of radius $R$, each eigenfunction is a
product of a spherical harmonic $Y_{\ell M }(\Omega)$ and a radial
eigenfunction ${\cal R}_{nl}$ which involves the spherical Bessel
\newpage\noindent
function $j_\ell$ of order $\ell$.  The one-electron energy is $ E_{n
\ell} = { \hbar^2 k_{n \ell}^2 / 2m}$ where $( k_{n \ell} R)$ is the
location of the $n$'th zero of $j_\ell$.

    At temperature $T=0$, the Fermi function $f(E_i)$ is replaced by the step
function $\theta (\mu - E_i)$ where $\mu$ is the
chemical potential. On evaluating the matrix element in Eq. (3),
we obtain
\begin{equation}
\begin{array}{l}
\varepsilon_2(\omega)= \displaystyle {{16 \pi e^2 \over m^2 R^5 \omega^4}
 \sum_{n_1 \ell_1 \atop n_2 \ell_2}}~
{1 \over 2} ~(\ell_1 \delta_{\ell_1,\ell_2+1}+\ell_2
\delta_{\ell_2,\ell_1+1})~ E_{n_1 \ell_1} E_{n_2 \ell_2} \\
\hbox{~~~~~~~~~~~~~~}
 \theta ( \mu - E_ {n_1 \ell_1})
\theta (E_{n_2 \ell_2} - \mu )~\delta
(\hbar \omega + E_{n_1 \ell_1} - E_{n_2 \ell_2}).
\end{array}
\end{equation}
In the limit $k_FR \rightarrow \infty$ where $k_F \equiv \sqrt{2\mu m
/ \hbar^2} $ is the Fermi wave vector, the sums over $(n_1,\ell_1)$ and
$(n_2,\ell_2)$ in Eq. (4) may be replaced by integrals. Though
several authors had earlier obtained this equation,
they had not evaluated the integral
correctly.\cite{kk,ry}
The correct procedure  involves using the Debye expansion
 for large order Bessel functions \cite{as}.
$\varepsilon_2(\omega)$ can be evaluated in closed form \cite {bs}:
\begin{equation}
\varepsilon_2(\omega) = {4 \over \pi} {e^2 \over \hbar \omega R}~ {1
\over \nu^3}~ G(\nu),
\end{equation}
where $ \nu \equiv \hbar \omega / \mu $ and
$ G(\nu)= g(\nu) - g(-\nu) $
with
\begin{equation}
g(\nu)= {(1+\nu)^{3 \over 2} \over 3 } + {\nu^2 (1+\nu)^{1 \over
2} \over 4} - {\nu^2 (2 + \nu ) \over 8} ~log \left\vert{ \sqrt {1+\nu} +1
\over \sqrt{1+\nu}-1}\right\vert.
\end{equation}

     In the limit of large radii, the real part $\varepsilon_1(\omega)$ of
the dielectric constant is approximately $ \varepsilon_1(\omega) = 1 -
{\omega_p^2 / \omega^2} .$
The absorption as a function of frequency is given by Eqs. (2), (5) and (6),
and the FWHM $\Gamma$ of the Mie resonance
is given by $2 \varepsilon_2 / | \partial \varepsilon_1/
\partial \omega |$ evaluated at $\omega_0.$
 The result is
\begin{equation}
\Gamma = {3 \over 4} {G(\nu_0) \over
\nu_0}  {v_F \over R},
\end{equation}
where $\nu_0 \equiv \hbar \omega_0 /\mu.$ In the limit $\nu
\rightarrow 0$, we see from Eq. (6) that the ratio $G(\nu )/\nu$ approaches
unity; the ratio is less than unity for nonzero values of $\nu$.

Clearly, the model used is very simple in many respects.
 Though it captures the crucial
finite-size aspect and predicts a $1/R$ dependence of the width, the
value of the slope is sensitive to a number of physical effects which
have not been included. As mentioned above, the experimentally
determined value
of the slope is sensitive to the outside matrix, suggesting that
outer surface effects are important, and the approximation of an
infinite square well may be too drastic. The correction coming from
a large but finite depth $V_0$ of the well has recently been computed
\cite {yr}
to leading order in $\mu/V_0$. Another extension that has been considered
\cite {yb} is to account for finite temperatures in leading order in
$T/\mu$. Both effects $(V_0 \ne \infty, T \ne 0)$ lead to a further broadening
of the line.
Finally, we mention the review \cite{apell}, which discusses
the interesting possibility that the dielectric constant itself may vary
in space, close to
\newpage\noindent
the surface of the metal particle. This has strong
effect on the charge density near the surface, and thus on the resonant
response.

\vskip 1cm
\noindent {\it 2.2. Resonances in Free Clusters}
\vskip .5cm

Experiments on samples of type (ii) have been performed on alkali
clusters with between two and hundreds of conduction electrons \cite
{sel,tiggs,brech}. The data for smaller clusters \cite{sel}
indicate that there is a strong
response over a relatively narrow frequency interval in the case of
magic numbers, and over a much broader frequency range in cases which
fall between magic numbers. In the latter case, the spectrum shows
splittings, which can be interpreted in terms of shape
deformations. For instance, a triaxial ellipsoid exhibits three
different resonance conditions and frequencies, corresponding to
different internal fields along the three axes. \cite{kresin}
 Even if we confine our
attention to magic clusters, however, there is no evidence for a
$\Gamma \sim 1/R$ law of the type seen in embedded samples. Nor is the
apparent conflict resolved by recent in-beam experiments on larger
free clusters of $K$ and $Li$.

Thermal excitation of surface fluctuations \cite {pab}, and consequent
 broadening of the resonance line, is a physical effect which may at
 least partially account for the observed differences of
 behaviour. This effect is expected to be absent in the embedded
 samples, as the surface of the metal particle is constrained not to
 move by the surrounding matrix. A controlled change of temperature
 (from $\sim 4 ^{\circ} K {\rm~~ to} \sim 300 ^{\circ} K$) produces
 virtually no change in the pattern of the resonant response
\cite {kreibig}. By
contrast, the response of type (ii) samples (in-beam clusters) shows a marked
dependence on the temperature $T$. A possible reason is that
 since the clusters
are in free flight, their shape is free to fluctuate. Thermal excitation
of such  surface-shape fluctuations is a possibility, given the relatively
high temperatures of the clusters, which reflect their process of formation.
An experimental  complication is that clusters of
widely different sizes are usually prepared
at different temperatures, making it difficult to disentangle effects
coming from variations of $R$ and $T$.

A simple argument yields the size dependence of $\Gamma$ that would
result from thermally excited shape fluctuations, if the temperature
were held constant. Consider deformations of shape around a
sphere, and suppose the shape changes occur on a time scale much
 longer than the probe time.
Let $\delta$ be a dimensionless measure of the
deformation. Assuming the volume is constant, the increase of surface
area is proportional to $\delta^2$, and the associated increase of surface
energy is $E= a \sigma R^2 \delta^2$
where $\sigma$ is the surface tension and $a$ is a constant of
order unity. At high enough $T$,
equipartition is valid, implying $<E> =  k_B T/2$, so that
\begin{equation}
    \delta_{RMS} \equiv <\delta^2>^{1/2} \sim \sqrt{ k_B T\over {\sigma}}
    { 1 \over R}.
\end{equation}
Since a value $\delta$ of the deformation causes a splitting of the Mie
resonance line into lines separated by $\omega_0 \delta$, the width
$\Gamma_{surf~fluc}$ resulting
from such surface fluctuation
\newpage\noindent
contributions is proportional to $\delta_{RMS}$:
\begin{equation}
\Gamma_{surf fluc} = c \omega_0
     \sqrt{ k_B T\over {\sigma}}
    { 1 \over R },
\end{equation}
where c is a constant. It is interesting that
an inverse-$R$ dependence shows up again, although the origin is quite
different
from the quantum size effect discussed in Section 2.1.
Part of the reason that this
$1/R$ dependence has not been seen in experiments may be that  the
temperatures used to prepare clusters of different sizes vary appreciably.

\vskip 1cm
\noindent {\bf 3. Giant Dipole Resonances in Nuclei}
\vskip 0.5cm
\noindent {\it 3.1. Cross Section Systematics}
\vskip 0.5cm

Photo-neutron cross sections have been measured in a large number of
nuclei; for a compilation of the data, see Ref. (21). For spherical nuclei,
the peak frequency $\omega_0$ is
known \cite{vdw} to exhibit a systematic empirical
 dependence on the mass number
$A$:
\begin{equation}
\omega_0 = 31.2 A^{-1/3} + 20.6 A^{-1/6}~{\rm MeV}.
\end{equation}
We shall discuss this dependence ({\it vis-\`a-vis} the
weak variation of $\omega_0$ with $A$ in metal particles) in Section 4.1. The
FWHM provides a simple, single characterization of the resonance
spectrum, and has been used earlier to extract global trends with
varying $A$, for heavy nuclei. For instance, Berg\`ere \cite {berg} and
Snover \cite {sno} have shown plots of the FWHM in the regions $A > 90$ and
$166 > A > 63$, respectively, including nuclei whose
resonance spectra exhibit split peaks.
These plots show that the FWHM exhibits
systematic oscillations in the ranges studied, with local minima near
spherical, near-magic nuclei.

We wanted to see whether the systematics observed earlier for the FWHM
of heavy nuclei \cite{berg,sno} persisted in lighter
nuclei as well. Accordingly, we examined the FWHM in about 120 nuclei
 ranging from
$^3 He$ to $^{239} Pu$, using primarily the cross-section data
compiled in Ref. (21).
(We re-examined the heavier
nuclei in order to have a uniform procedure for all $A$.) For those
cases where the data follows a curve with a single peak, it was
straightforward to determine the FWHM. For cases with two or more
(closely overlapping) peaks, we found the FWHM by drawing a smooth
curve with a single maximum through the data points, trying to ensure
that the areas under the smooth curve and the experimental data were
nearly equal.  Those nuclei where the data seem incomplete
($^3H,~^{19}F$) or have too much structure ($^{14}C,
{}~^{18}O,~^{24,26}Mg$) were ignored. Results are displayed in Fig.
1. For light nuclei (see inset in Fig. 1), we also estimated the
errors in the FWHMs, arising from (a) the existence of more than one
data set in some cases, and (b) the inherent uncertainty in extracting
the FWHM by our procedure. In the range $A > 90$, our values agree
well with those of Berg\'ere \cite {berg}.
\newpage

Examination of the results in Fig. 1 for light nuclei ($A< 50$) (see
the inset) shows that the FWHM continues to display local minima at,
by and large, the magic numbers. The rapid oscillations of the FWHM
versus $A$ are due to the relative crowding in of magic numbers for
small $A$. With the sole exception of $^{28}Si$, all the minima occur
at or near the magic numbers.
(Although $^{28}Si$ is not a magic nucleus, it displays
behaviour similar to a magic nucleus in at least one other context:
the plot of nuclear electric quadrupole moment vs. $Z$ or $N$ passes
through a zero near $^{28}Si$, indicating a prolate to oblate
transition.\cite {pb})
Conversely, each magic number
has a corresponding minimum, with the possible exception of $N = 40$
($A = 72$), where there is a hint of a local minimum, but the data
does not allow us to draw a firm conclusion. In any case, 40 is known
to be a weak magic number.
\begin{figure}[H]
\vspace{13 cm}
\fcaption{Full width at half maximum $\Gamma$ of the
total photoneutron cross section data, Ref. (21), versus the nucleon
number $A$. Note the systematic modulations, with minima at the
proton $(Z)$ or neutron $(N)$ magic numbers. The curve is drawn as a
guide to the eye. Dashed lines indicate regions of sparse or
nonexistent data. The inset shows the region $A\le 45$ in greater
detail.}
\end{figure}
\newpage

The systematic oscillations in the region $A<50$ in Fig. 1 are
statistically significant. As is evident from the inset, the error in
the FWHM is less than 1 MeV in almost all cases, and is generally much
smaller.  The amplitude of oscillations, on the other hand, is at
least 5-6 MeV ({\it e.g.}, $^3He$ to $^4He$, or $^{40}Ca$ to
$^{45}Sc$), and is sometimes as large as 14 MeV ({\it e.g.}, $^9Be$ to
$^{14}N$). The oscillations are as systematic and as pronounced as
those for large $A$, the only difference being that there are fewer
points per oscillation.

That the photo-response of a nucleus even as light as $He$ can be
thought of in the same terms as that of heavier nuclei may seem
surprising, but the very fact that the FWHMs for light nuclei fit in
well with the systematics across the periodic table provides an {\it a
posteriori} justification for the use of the FWHM even for $A<50$.

An interesting feature of Fig. 1 is the overall downward trend of
the oscillatory curve, evident if, for instance, we focus on points in
the lower envelope of the curve. These points correspond mostly to
spherical, magic nuclei. The observed downward trend is further
discussed in Section 4.2.

In the rest of this section we present schematic theoretical
considerations aimed at understanding the empirical systematics,
in contrast to more customary detailed theoretical studies of the
width for individual nuclei.

\vskip 1cm

\noindent {\it 3.2. Resonance Widths --- Overall Trend}
\vskip 0.5cm

One may distinguish between two types
of contributions to the FWHM. Firstly, there is the intrinsic width of the
resonance which comes from the finite lifetime of the collective mode,
and which is present in all cases. The intrinsic width itself receives
contributions from a variety of physical mechanisms to be discussed
below. This is the only contribution to the FWHM in spherical nuclei.
Secondly, in nonspherical nuclei, static deformations in shape can
lead to two distinct resonance frequencies, corresponding to a
splitting of the line. In such cases, the FWHM receives additional
contributions.

The intrinsic width $\Gamma_i$ can be written as the sum of three
terms (see, {\it e.g.}, Ref. (7)).
\begin{equation}
\Gamma_i =\Delta\Gamma
+ \Gamma^{\uparrow} + \Gamma^{\downarrow},
\end{equation}
reflecting contributions from distinct physical effects. The
fragmentation width $\Delta\Gamma$ corresponds to the fact that the
collective $(1p-1h)$ state which is the doorway state for the GDR, is
not a single state, but is in most cases already appreciably
fragmented. This effect (mean-field damping or one-body friction) is
the finite nucleus analogue of Landau damping in a bulk medium. It
occurs due to the scattering of the nucleons from the `wall' or
`surface' of the self-consistent mean field potential. This is
the physical effect which was accounted for by the quantum calculation
 of the resonance
width of a metal particle, discussed in Section 2.1 above.
The second term
$\Gamma^{\uparrow}$ is the escape or decay width corresponding to the
direct coupling of the $(1p-1h)$ doorway state to the continuum,
giving rise to its decay into a free nucleon and an $(A-1)$
\newpage\noindent
nucleus.
Finally, the spreading width $\Gamma^{\downarrow}$ is due to the
coupling of the $(1p-1h)$ doorway state to more complicated $(2p-2h)$
states of the nucleus, the transition occurring on account of genuine
two-body effects (collisional damping or two-body friction).

Let us see how each contribution to Eq. (11) is expected to vary with
radius $R$. Our arguments are schematic, and aimed at establishing the
general, systematic trend with size.

The $R$-dependence of the first two terms may be estimated using a
simple argument based on estimating the frequency of collisions with
the surface. Such an argument has been used successfully \cite {kreibig}
to estimate $\Delta\Gamma$ in metal particles; the estimate
agrees with the result of the quantum calculation presented in
Section 2.1.
The idea is that individual fermions moving with Fermi velocity $v_F$
hit the wall with mean time $\sim R/v_F$; the inverse time $\sim
v_F/R$ then determines the contribution to the width arising from wall
effects --- both for $\Delta\Gamma$ and $\Gamma^{\uparrow}$.  The
subject of one-body dissipation has also been discussed in
the nuclear physics literature. \cite {brink}
The spreading width $\Gamma^{\downarrow}$, on the other hand, arises from
two-particle collisions. We expect $\Gamma^{\downarrow}$ to vary
smoothly with energy and nuclear size for magic cases, since the
collisional mean free path $\Lambda$ shows similar smooth variations
(see, {\it e.g.}, Ref. (26)). The average time between two
collisions is $\sim\Lambda/v_F$ and hence $\Gamma^{\downarrow}$ is
expected to be $\sim v_F/\Lambda$. In the limit $R\rightarrow\infty$,
this is the only contribution.

The total intrinsic width is thus expected to be of the same form as
Eq. (1) with $\Gamma_\infty$ being the spreading width in the
$R\rightarrow\infty$
limit.

\vskip 1cm
\noindent {\it 3.3. Resonance Widths --- Shell Effects}
\vskip 0.5cm

In Section 3.2, we discussed only the monotonic variation of the FWHM that
obtains for spherical nuclei. As we see from Fig. 1, when we
consider $all$ nuclei, superimposed on this monotonic variation, there
are striking and strong shell-structure-linked oscillations in the
FWHM. We discuss the origin of these
oscillations in two broad representative regions, namely $150<A<190$
and $80<A<150$.

In the range $150<A<190$, Dietrich and Berman \cite {db} have fitted
two-component Lorentz curves to the photoneutron cross section
data. This indicates a splitting of the line, due to deformation of
the nucleus. We denote the two resonance energies
by $\omega_{01}$ and $\omega_{02}$, with $\omega_{01}<\omega_{02}$,
and the corresponding widths by $\Gamma_1$ and $\Gamma_2$. For a
spheroidal deformation, Eq. (10) leads us to expect that $\omega_{01}$ and
$\omega_{02}$ correspond to oscillations along the semimajor and
semiminor axes respectively. Equation (1) then implies that $\Gamma_1
< \Gamma_2$.  This is indeed found to be true, for $150<A<190$, for
the values of the widths tabulated in Ref. (21). In
fact, with the exceptions of $^{55}Mn$ and $^{63}Cu$, this is true for
all the nuclei listed there. This provides additional evidence for
the overall decrease of the intrinsic width with increasing
radius. The increase of the FWHM
away from the spherical cases can be ascribed, at least partially, to
the fact that deformations produce a splitting of the line, and also
cause $\Gamma_2>\Gamma( >
\Gamma_1)$,
\newpage\noindent
 where $\Gamma$ would be the width if the nucleus were
undeformed. Since deformations of the shape follow shell-structure
systematics, with smallest deformations close to the magic numbers, so
does the FWHM \cite {oka}.

In the region $80<A<150$, Dietrich and Berman \cite {db} have fitted
one-component Lorentz curves to the photoneutron cross section data
(with the exceptions of $^{127}I$ and $^{148}Nd$). As is clear from
Fig. 1, oscillations of the FWHM versus $A$ in this region are as
prominent as those for $150<A<190$. Thus, even when the line is
unsplit, the width of the best-fit Lorentzian oscillates as a function
of $A$, with minima at the magic numbers. This indicates that the
intrinsic width $\Gamma_i$ can itself show shell-structure-linked
oscillations. Berg\`ere \cite{berg} has correlated the FWHM in
this region with the ratio $E(4^+)/E(2^+)$, which characterizes the
`softness' of nuclei. Here $E(J^+)$ is the energy of the first $J^+$
state in the nuclear spectrum.

\vskip 1cm
\noindent {\bf 4. Similarities and Differences between Nuclei and Metal
Particles}
\vskip 0.5cm
\noindent {\it 4.1. Resonance Frequency}
\vskip 0.5cm

The size dependence of the natural frequency of vibration $\omega_0$
can be deduced by using a simple classical picture of the collective
mode.

 In the metal particle, the restoring force arises from the electric
field produced by layers of opposite charges on diametrically opposite
sides, and acts on each of the $A$ conduction electrons in the
particle. The oscillator frequency $\omega_0$ is given by the square
root of the ratio of the total restoring force per unit displacement
to the mass involved. Since both force and mass are proportional to
$A$, the frequency $\omega_0$ is roughly size independent.

In the nucleus, on the other hand, the restoring force
arises from short-range strong interactions amongst nucleons. In a
hydrodynamic description, it is modelled by the surface or volume
symmetry energy terms in the semiempirical mass formulas. In the
Goldhaber-Teller model \cite {gt}, the collective
state corresponds to the motion of the proton cloud through the
neutron cloud without mutual distortion. The restoring force is
proportional to $A^{2/3}$ and the mass parameter is proportional to
$A$.  Hence, $\omega_0 \sim A^{-1/6}$. In the Steinwedel-Jensen model
\cite {sj}, on the other hand, the relative
proton-neutron density changes in such a way as to maintain constant
overall density throughout. The restoring force per unit mass is
proportional to $R^{-2}$, and hence $\omega_0 \sim A^{-1/3}$.
(See Eq. (10).)

\vskip 1cm
\noindent {\it 4.2. Resonance Width}
\vskip 0.5cm

We wanted to see if Eq. (1), which holds for embedded metal particles,
 also describes the downward trend of $\Gamma$ in nuclei with
increasing $A$, evident in Fig. 1. A similar $1/R$ dependence has
been discussed earlier \cite {myers} for nuclei in the
range $A > 50$.
\newpage\noindent
 On dividing across by the Fermi energy $\epsilon_F$,
we see that Eq. (1) predicts that $\Gamma_{av}/\epsilon_F$ is a linear
function of $(k_F R)^{-1}$, where $k_F$ is the Fermi wave vector.
Interestingly, on using these dimensionless scaled variables, we can
directly compare the $Ag$-particle and nuclear data (Fig. 2) which
in absolute terms differ by six orders of magnitude

\begin{figure}[H]
\vspace{14.8 cm}
\fcaption{
$\Gamma/\epsilon_F$ versus $(k_F {\tilde R})^{-1}$
for embedded metal particles and nuclei.  The dashed and dotted lines
are best fits for $Ag/Ar$ ($+$, Refs. (9,31)) and $Ag/Ne$
Ref. (9). The nuclei shown here are singly $(\bullet)$ or doubly
$(\circ)$ magic nuclei from the lower envelope of the oscillating
curve in Fig. 1. The solid straight line is the best fit to this
data set. The dot-dashed line indicates the `average' trend of the
oscillatory curve in Fig. 1. Note the similarities between the
scaled nuclear and particle data despite the fact that the two data
sets differ by six orders of magnitude in energy and five orders of
magnitude in length.  For $A \ge 90$, not all error bars are shown;
nuclei in the same cluster have roughly similar error bars.}
\end{figure}
\newpage

\noindent{in energy and five
orders of magnitude in length. We used the values $\epsilon_F = 38$
MeV and $k_F = 1.36$ fm$^{-1}$ for nuclei, and $\epsilon_F = 5.49$ eV
and $k_F = 1.20 A^{-1}$ for $Ag$ particles. We have chosen to plot
data for $Ag$ particles in argon and neon matrices as interactions
with surrounding inert gas atoms are likely to be minimal, and a large
range of sizes has been studied for $Ag/Ar$. \cite{kg,charle}
We have used RMS radii $\tilde R$, as these are well
determined for nuclei; for $Ag$ particles, we took $\tilde R$ to be
given by $\sqrt {3/5}$ times the quoted radii. Since $\Gamma$
oscillates as a function of size in nuclei, and we are interested in
displaying the overall downward trend, we have replotted points
corresponding to singly or doubly magic nuclei from the lower envelope
of the curve in Fig. 1; the line marked `magic' is the best fit line
through these points. Thus these points are consistent with a linear
dependence on $(k_F \tilde R)^{-1}$, though other monotonic variations
with $\tilde R$ cannot be ruled out. We also examined the average
downward trend of the oscillatory curve in Fig. 1, and found that it
could also be fit to a linear dependence. The slope of the average
line (marked `average' in Fig. 2) is larger than that of the solid
line, and is comparable to the slopes of the dashed and dotted lines.
Thus (1) holds to a good approximation for nuclei also. In particular,
it is interesting to see how well the doubly magic nuclei $^4He,~
^{16}O,~ ^{40}Ca,~ ^{90}Zr,$ and $^{208}Pb$ follow a straight line.}

Of course, there are also some differences between nuclei and metal
particles. Consider the limit $R\rightarrow\infty$,
corresponding to nuclear matter or the bulk metal. From Fig. 2, we
see that if the straight lines are extrapolated towards $(k_F{\tilde
R})^{-1}=0$, the resulting intercept on the $\Gamma/\epsilon_F$ axis
is much larger for nuclei than for metal particles. This is presumably
because $k_F \Lambda$ is much smaller in nuclear matter
\cite {wam} than in bulk metals at room temperature \cite {am}
reflecting the greater effect of collisions in the
former case. As a result,
the limiting contribution $\Gamma_\infty$
constitutes a substantial fraction of the total width for spherical
nuclei (for $^{208}Pb$ it is about $50\%$), while
for metal particles of comparable $A$, the contribution
$\Gamma_\infty$ is a much smaller fraction of the width.

\vskip 1cm
\noindent {\bf 5. Conclusion}
\vskip 0.5cm

We conclude by recapitulating the main points of this paper.

The $1/R$ law (Eq. (1)) describing the observed broadening of the
width of the Mie resonance in embedded metal particles is captured by
a simple model of electrons in a spherical well. The quantum
mechanical response can be computed analytically when the well is
infinitely deep, but the model is too simple to correctly predict
experimental values of the slope, which are sensitive to the
surrounding medium. Smaller metal clusters in free flight exhibit a
more complicated behaviour, perhaps because of the thermal excitation
of surface fluctuations. At high temperatures, the resulting
broadening is proportional to ${\sqrt T}/R$.

In nuclei, the FWHM of the total photoneutron cross section shows
shell-structure-linked oscillations as a function of $A$ even for
$3<A<50$. Disregarding oscillations, for instance by focusing on magic
nuclei, the FWHM generally decreases with
\newpage\noindent
 increasing $A$ approximately
as $A^{-1/3} \sim 1/R$. While a complete theory has not been presented
here, we have given a schematic theoretical description which allows
one to understand at least the principal trends.

Striking similarities are seen when the FWHMs for nuclei are compared
with photoabsorption FWHMs in embedded metal particles, after proper
rescaling of the energies and lengths (Fig. (2)).

\vskip 1cm
\noindent {\bf 6. Acknowledgements}
\vskip 0.5cm

MB acknowledges fruitful interactions and discussions with
V. Subrahmanyam.

\vskip 1cm

After this paper was written, we received a copy of the monograph by
Bertsch and Broglia \cite {bebr} which deals with dipolar oscillations
in metal clusters and nuclei, and bears on many of the points
discussed in this paper.

\vskip 1cm

\end{document}